# Low temperature synthesis and pressure induced insulator-metal transition of the newly found NdFeAsO$_{0.75}$

Qingping Ding, Yangguang Shi, Hongbo Huang, Shaolong Tang, Shaoguang Yang*

National Laboratory of Microstructures, Nanjing University, Nanjing 210093, P. R. China

E-mail: sgyang@nju.edu.cn

From the discovery of superconductivity with transition temperature ($T_c$) of 26 K in LaO$_{1-x}$F$_x$FeAs system,[1] iron-based layered compound ReFeAsO attracted much attention in the scientific community. More and more efforts have been devoted in the study of this kind of materials. The transition temperature has soon been enhanced a lot,[2-7] now the reported highest $T_c$ is 56.3 K.[8] But much work remains in this field, for example, synthesis of such iron-based compounds in an easy way remains a serious problem. Quebe P. et al reported a synthesis method at low temperature with KCl/NaCl as mineralizer,[9] but it is difficult to avoid unwanted element contamination with this method. The synthesis temperature in all other reported methods was higher than 1150 $^{o}$C. [1-8,10,11] This makes it difficult in the preparation of such kind of materials, especially for the strength of the quartz tube for sealing the vacuum at high temperature. So synthesis of the newly found tetragonal phase iron-based compounds at relatively low temperature is very important currently. Here we report an illustration of synthesis of iron-based compound NdFeAsO$_{0.75}$ at a temperature of 900 $^{o}$C. This synthesis temperature is the lowest among all the reported methods without the help of any mineralizer. In the materials research region, high pressure is often used as an effective parameter to tailor the property of materials. In iron-based superconductors, it has been confirmed that pressure can either enhance or suppress the $T_c$ by both theories and experiments.[12-15] By applying high pressure to our $NdFeAsO_{0.75}$ sample prepared at 900 °C, pressure induced insulator to metal phase transformation in $NdFeAsO_{0.75}$ is reported. To our knowledge, this is the first experiment of insulator-metal transition in iron-based layered compound ReFeAsO.

In our present work, the polycrystalline sample was prepared by the conventional solid state reaction. NdAs, Fe, $Fe_2O_3$ were used as starting materials and weighed according to the chemical stoichiometry of $NdFeAsO_{0.75}$. NdAs was obtained by reacting Nd chips with As pieces at 600 °C for 5 hours and then 900 °C for 10 hours. The raw materials were thoroughly grinded and pressed into pellets. Then pellets were sealed in an evacuated quartz tube, and annealed at 900 °C for 36 hours. The prepared sample was further treated under a pressure of 6 GPa and temperature of 1300 °C for 2 hours.

The structure of the samples prepared at 900 °C and after high pressure were characterized by a powder X-ray diffraction (XRD) method with Cu Kα radiation (λ=1.5418 Å) in the 2θ range of 20−80 degree with the step of 0.02 degree at room temperature. Scanning electronic microscopy (SEM) was used to characterize the morphologies of the samples. The SEM analysis was performed on a Philips XL30 microscope operated at 20.0 KV. To realize the temperature dependence of resistivity, standard 4-probe dc resistivity measurements were preformed from 300 K down to 2 K in a Physical Property Measurement System (PPMS) of Quantum Design Company.

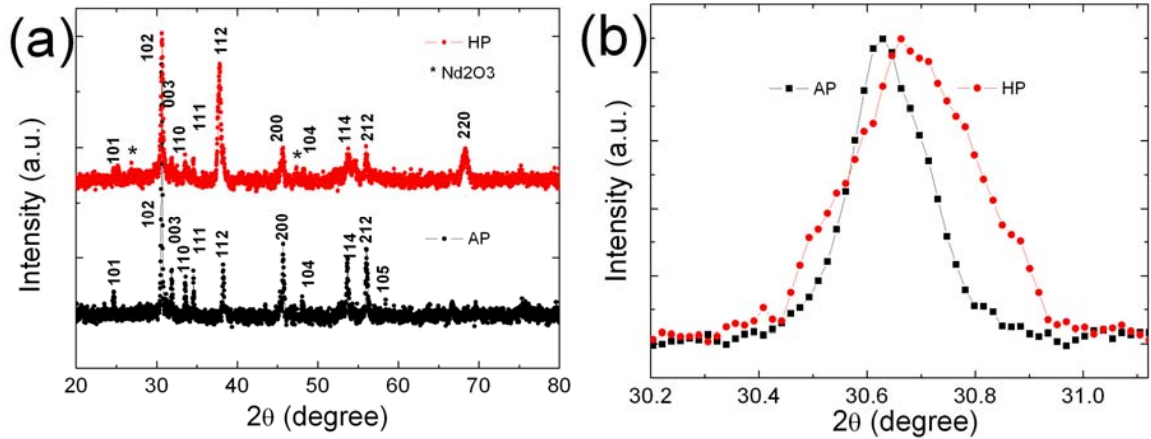

***Figure 1.*** (a) X-ray diffraction patterns of NdFeAsO$_{0.75}$ before and after high pressure treatment, most peaks could be well indexed on the basis of tetragonal ZrCuSiAs-type structure with the space group P4/nmm. AP: ambient pressure, without high pressure treatment; HP: after high pressure treatment. (b) Magnification of peak (102) in figure 1a. Peak (102) shifted from 30.629 to 30.663 degree, which means that the distance between two neighboring (102) was compressed after the high pressure treatment.

The XRD patterns for the prepared samples are shown in figure 1a, most peaks could be well indexed on the basis of tetragonal ZrCuSiAs-type structure with the space group P4/nmm. Almost pure phase was achieved for the samples before high pressure treatment. Two weak peaks (maked with *) assigned to Nd$_2$O$_3$ can be observed in the sample after high pressure treatment, which shows that very small impurity was formed in the high pressure process. Part of the highest XRD peak (102) was magnified as shown in figure 1b. From this figure, it can be found that the (102) peak moved from 30.629 to 30.663 degree. It means that the distance between two neighboring (102) was compressed after the high pressure treatment.

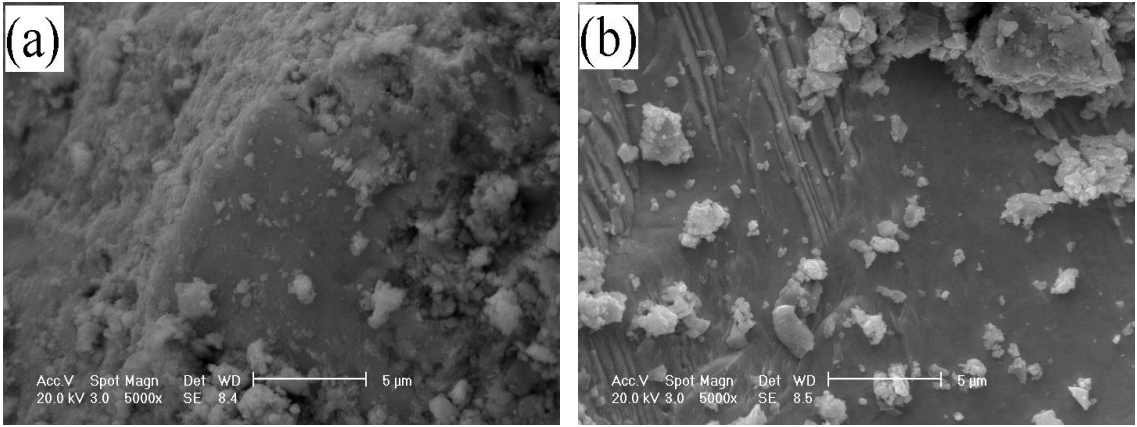

*Figure 2.* SEM images of NdFeAsO$_{0.75}$ before (a) and after (b) high pressure treatment. The sample after high pressure treatment shows better compactness and layered structure feature.

SEM images of NdFeAsO$_{0.75}$ samples are presented in figure 2. Figure 2a and figure 2b are typical images of the sample before and after high pressure treatment. Compared with the sample prepared at 900 $^{o}$C, the sample after high pressure treatment shows better compactness and layered structure feature. These two images were recorded under the same condition, but the definition of figure 2b is better than that of figure 2a. This means the sample after high pressure treatment shows better electrical transport ability than the untreated sample.

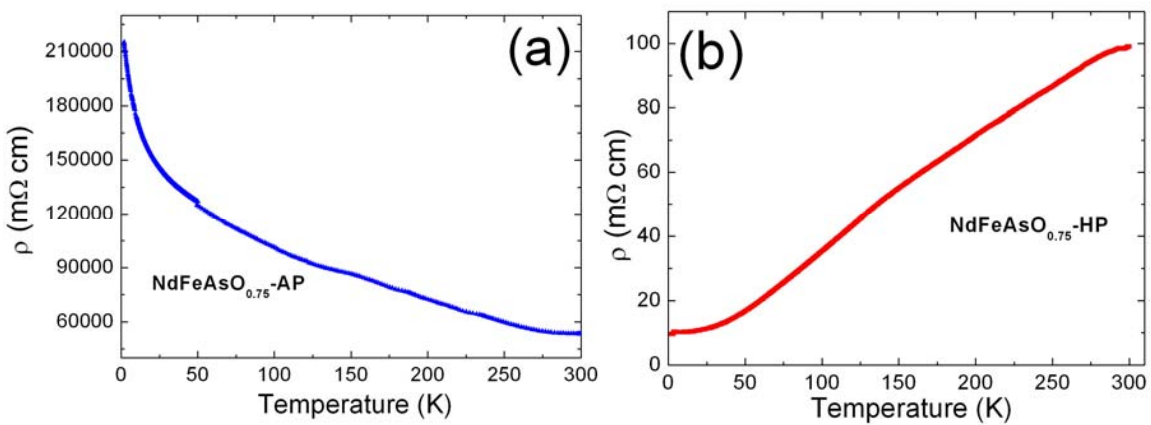

*Figure 3.* R-T measurement results of NdFeAsO$_{0.75}$ before (a) and after (b) high pressure treatment. The resistivity has decreased 3-4 orders in the measured temperature range, the sample has been changed from insulator-like phase to metal-like phase.

Figure 3a shows the temperature dependence of the resistivity of the 900 $^{o}$C synthesis sample. From this figure, it can be found that the resistivity decreases with the temperature increase and the dc resistivity value is rather high. Figure 3b shows the dc resistivity dependence of the temperature of the sample treated after high pressure. The resistivity increases with the temperature, and the resistivity is much smaller than before high pressure treatment at the same temperature. Compared the resistivity results of the two samples, a very clear difference between them can be found. The resisitity has decreased 500 times (at 300 K) to 21000 times (at 2K), and the sample has been changed from insulator-like phase to metal-like phase after the high pressure treatment.

From the XRD measurements, it can be concluded that both samples were almost pure tetragonal phase NdFeAsO$_{0.75}$. The electrical transport property mainly comes from the tetragonal phase NdFeAsO$_{0.75}$. Although very small amount of Nd$_2$O$_3$ was observed in the high pressure treated sample, this little Nd$_2$O$_3$ should not influence the tendency of the main phase of the sample. As we know, by modifying the chemical composition or the lattice parameters the electron correlation strength can be controlled while essentially keeping the original lattice structure unchanged.[16] By control of the transfer interaction or the one-electron bandwidth, control of electron correlation strength usually can be achieved. Pressure is often used as an effective parameter to tailor the one-electron bandwidth. Generally speaking, applying a pressure decreases the interatomic distance and thus increases the transfer interaction. This may be the reason of great decrease of the electric resistivity and the insulator-metal transition.

To summarize, low temperature synthesis method has been successfully performed in the synthesis of tetragonal phase NdFeAsO$_{0.75}$. The synthesis method may be applied for other rare earth substitution iron based tetragonal phase materials, ReFeAsO1-xFx or ReFeAsO1-x (RE = La, Ce, Pr, Nd, Sm, Gd) compounds. XRD measurements illustrate that almost pure phase tetragonal phase NdFeAsO$_{0.75}$ for both two samples before and after high pressure treatment. Temperature dependence of resistivity shows that its electrical transport property has changed from insulator-like phase to metal-like phase. Although further theoretical research such as first principle calculation is needed to quantify the great decrease of resistivity and this insulator-metal transition. The present results add a new and important feature into the transport property of iron-based layered compound ReFeAsO. More phase transitions are expected via applying high temperature high pressure process treatment.

**Aknowledgement.** This work was supported by the NSFC (10774068), NCET(07-0430) and 973 Program(2006CB921800).